\documentclass[aps,prc,groupedaddress,twocolumn,floatfix]{revtex4-2}

\usepackage{float}
\usepackage{graphicx}
\usepackage{amsmath}
\usepackage{lipsum}
\usepackage{amssymb}
\usepackage[colorlinks=true, allcolors=blue]{hyperref}

\def\ba{\begin{eqnarray}}
\def\ea{\end{eqnarray}}
\def\be{\begin{equation}}
\def\ee{\end{equation}}

\begin{document}

\title{Bottomonium suppression at RHIC and LHC in an open quantum system approach}
\author{Michael Strickland}
\author{Sabin Thapa}
\affiliation{Kent State University, Department of Physics, Kent, OH 44242 USA}

\begin{abstract}
We present potential non-relativistic quantum chromodynamics (pNRQCD) predictions for bottomonium suppression in $\sqrt{s_{NN}}$ = 200 GeV, 2.76 TeV, and 5.02 TeV heavy-ion collisions using an open quantum systems (OQS) description of the reduced heavy-quark anti-quark density matrix.  Compared to prior OQS+pNRQCD studies we include the rapidity dependence of bottomonium production and evolution, allowing for a fully 3-dimensional description of bottomonium trajectories in the quark-gluon plasma.  The underlying formalism used to compute the ground and excited state survival probabilities is based on a Lindblad equation that is accurate to next-to-leading order (NLO) in the binding energy over temperature.  For the background evolution, we make use of a 3+1D viscous hydrodynamics code which reproduces soft hadron observables at all three collision energies.  We find good agreement between NLO OQS+pNRQCD predictions and data taken at LHC energies, however, at RHIC energies, there is tension with recent bottomonium suppression measurements by the STAR collaboration.
\end{abstract}


\maketitle

\section{Introduction}

In the 1980s Matsui and Satz proposed heavy quarkonium suppression as a potential indicator of quark-gluon plasma (QGP) formation~\cite{Matsui:1986dk}.
Since then, the study of quarkonium suppression has become a crucial component of the heavy-ion research programs at various experimental facilities.
For recent investigations conducted at LHC and RHIC experiments see Refs.~\cite{STAR:2013kwk,PHENIX:2014tbe,ALICE:2014wnc,STAR:2016pof,CMS:2016rpc,CMS:2017ycw,Sirunyan:2018nsz,CMS:2018zza,ALICE:2019pox,Acharya:2020kls,CMS:2020efs,ATLAS:2022xso,STAR:2022rpk,CMS:2023lfu}.
The conjecture by Matsui and Satz in \cite{Matsui:1986dk} was that the chromoelectric fields in the medium are screened at distances proportional to the inverse of the Debye mass, $m_D \propto T$, where $T$ is temperature of the system.  In the Matsui-Satz picture, in-medium screening results in the dissociation of quarkonium, with quarkonium states having different radii dissociating at different temperatures, which could provide a quarkonium-production based QGP thermometer.  

In recent years, however, the Matsui-Satz screening paradigm has been challenged by first-principles calculations that demonstrated that the potential experienced by quarks/antiquarks in the QGP possesses both real and imaginary contributions \cite{Laine:2006ns,Beraudo:2007ky,Escobedo:2008sy,Brambilla:2008cx,Dumitru:2009fy,Brambilla:2010vq,Brambilla:2011sg,Brambilla:2013dpa}, with the imaginary parts being large enough that they result in decays of in-medium states on the fm/c timescale.  As a result, it is the imaginary part of the potential, rather than screening effects, which dominates in-medium heavy-quarkonium suppression.  There exist various theoretical approaches to compute quarkonium suppression including the effect of in-medium widths, which include quantum \cite{Laine:2006ns,Beraudo:2007ky,Brambilla:2008cx,Escobedo:2008sy,Brambilla:2010vq,Akamatsu:2011se,Strickland:2011mw,Strickland:2011aa,Akamatsu:2014qsa,Blaizot:2015hya,Krouppa:2015yoa,Katz:2015qja,Brambilla:2016wgg,Krouppa:2016jcl,Blaizot:2017ypk,Brambilla:2017zei,Krouppa:2017jlg,Yao:2018nmy,Brambilla:2019tpt,Rothkopf:2019ipj,Islam:2020gdv,Islam:2020bnp,Brambilla:2020qwo,Akamatsu:2020ypb,Yao:2020xzw,Yao:2020eqy,Brambilla:2021wkt,Omar:2021kra,Blaizot:2021xqa,Yao:2021lus,Brambilla:2022ynh,Alalawi:2022gul,Brambilla:2023hkw,Strickland:2023nfm} and transport approaches \cite{Grandchamp:2005yw,Rapp:2008tf,Du:2017hss,Yao:2018nmy,Du:2019tjf,Hatwar:2020esf,Yao:2020xzw,Yao:2020eqy}.

Using open quantum system (OQS) methods within the potential non-relativistic QCD effective theory Refs.~\cite{Brambilla:2016wgg,Brambilla:2017zei} demonstrated that the in-medium quantum evolution of heavy quarkonium depends on the heavy-quarkonium momentum diffusion coefficient $\hat{\kappa}$ and its dispersive counterpart $\hat{\gamma}$, with both being defined in terms of nonperturbative correlators of chromoelectric fields.  The quantum master equations governing the evolution of the heavy-quark reduced density matrix were obtained using OQS methods in which one integrates out the light medium degrees of freedom to obtain evolution equations for the heavy quarkonium reduced matrix.  For application to heavy-ion collision phenomenology one considers a strongly-coupled quark-gluon plasma obeying the scale hierarchy $M \gg 1/a_0 \gg m_D,T \gg E$, where $M$ is the heavy-quark mass, $a_0$ is the vacuum 1S Bohr radius, $m_D$ is the Debye mass, $T$ is the temperature, and $E$ is the binding energy.

Within this hierarchy one obtains a Markovian quantum master equation due to the fact that the medium relaxation time $\tau_{\rm med} \sim 1/T$ is much smaller than both the timescale for internal transitions $\tau_{\rm int} \sim 1/E$ and the probe relaxation time $\tau_{\rm probe} \gg 1/T$.  In this case one can obtain a Markovian quantum master equation of Lindblad form~\cite{Brambilla:2016wgg,Brambilla:2017zei,Rothkopf:2019ipj,Akamatsu:2020ypb,Yao:2021lus}.  The resulting Lindblad equation was solved numerically using the quantum trajectories method and comparisons of the phenomenological predictions with data are quite favorable as a function of both the number of participants $N_{\rm part}$ and transverse momentum $p_T$~\cite{Brambilla:2020qwo,Brambilla:2021wkt, Brambilla:2022ynh,Brambilla:2023hkw}.  The OQS+pNRQCD framework was applied first at leading-order in the expansion in binding energy over temperature ($E/T$) in Refs.~\cite{Brambilla:2020qwo,Brambilla:2021wkt} and then extended to next-to-leading-order (NLO) in $E/T$ in Ref.~\cite{Brambilla:2022ynh}.  

In this paper, we apply the same NLO framework derived in Ref.~\cite{Brambilla:2022ynh} and extend the phenomenological study presented in Ref.~\cite{Brambilla:2023hkw} to include predictions for the momentum rapidity ($y$) dependence of bottomonium energies.  Unlike prior works, the Lindblad equation solution is now fully coupled to the 3+1D hydrodynamics background due to the fact that the bottomonium states can have non-zero momentum rapidity~\footnote{Previous phenomenological applications of the OQS+pNRQCD framework explicitly assumed that $y=0$ due to computational limitations.}.  Herein, we consider three different collision energies, 5.02 TeV, 2.76 TeV, and 200 GeV in order to assess how well the approach works.  We find that including the full 3D dynamics improves agreement with available experimental data at the two highest energies.  At lower beam energies, we find poorer agreement with experimental data, independent of whether the rapidity dependence of bottomonium production is included or not.

The structure of our paper is as follows.  In Sec.~\ref{sec:methodology} we briefly review the NLO Lindblad equation for the evolution of the heavy-quarkonium reduced density matrix.  In this section, we also present our method for numerical solution of the Lindblad equation.  In Sec.~\ref{sec:hydro} we provide the details of the hydrodynamical evolution used for the background evolution.  In Sec.~\ref{sec:feeddown} we specify the manner in which late-time feed down of excited bottomonium states is taken into account.  In Sec.~\ref{sec:results} we present our results obtained at all three collision energies.  Finally, in Sec.~\ref{sec:conclusions} we present our conclusions and an outlook for the future.

\section{Methodology}
\label{sec:methodology}

In order to describe the evolution of pairs of heavy-quarkonium quark and anti-quarks we numerically solve the Lindblad equation for the evolution of the heavy-quarkonium reduced density matrix.  The NLO evolution equations were originally obtained in \cite{Brambilla:2022ynh}.  Below, we first provide a summary of the results found therein.  We then describe the numerical method used.  We note, importantly, that similar to \cite{Brambilla:2023hkw}, we include the effect of quantum jumps on the system's evolution.

\subsection{NLO pNRQCD+OQS}

We assume that the evolution of heavy quarkonium inside the quark-gluon plasma (QGP) can be described by a non-equilibrium master equation which can be derived using pNRQCD and OQS. As mentioned above, we make use of the following hierarchy of scales $M \gg 1/a_{0} \gg \pi T \sim m_{D} \gg E$, where $M$ is the heavy-quark mass, $a_{0}$ is the Bohr radius of the quarkonium, $T$ is the temperature of the medium, $m_{D} \sim gT$ is the Debye mass, and $E$ is the binding energy of the quarkonium state \cite{Brambilla:2016wgg,Brambilla:2017zei}. At NLO in the binding energy over temperature, the resulting Lindblad equation can be written as \cite{Brambilla:2022ynh}
\begin{eqnarray}
\frac{d\rho(t)}{dt} &=& -i \left[ H, \rho(t) \right] \nonumber \\ && + \sum_{n=0}^1
\left( C_{i}^{n} \rho(t) C^{n\dagger}_{i} - \frac{1}{2} \left\{ C^{n \dagger}_{i} C_{i}^{n}, \rho(t) \right\} \right),
\label{eq:Lindblad}
\end{eqnarray}
with reduced density matrix and Hamiltonian given by
\begin{equation}
    \rho(t) = \begin{pmatrix} \rho_{s}(t) & 0 \\ 0 & \rho_{o}(t) \end{pmatrix} \, ,
\end{equation}
and
\begin{equation}
	H = \begin{pmatrix} h_{s} + \text{Im}(\Sigma_{s}) & 0 \\ 0 & h_{o} + \text{Im}(\Sigma_{o}) \end{pmatrix} .
\end{equation}
At NLO, the singlet and octet self-energies are given by
\begin{eqnarray}
\label{eq:self_energies_im}
	\text{Im}\left( \Sigma_{s} \right) &=& \frac{r^{2}}{2} \gamma +\frac{\kappa}{4MT} \{r_{i}, p_{i}\} \, , \\
	\text{Im}\left( \Sigma_{o} \right) &=& \frac{N_{c}^{2}-2}{2(N_{c}^{2}-1)} \left( \frac{r^{2}}{2} \gamma +\frac{\kappa}{4MT} \{r_{i}, p_{i}\} \right) \, .
\end{eqnarray}
The collapse operators are
\begin{align}
	&\begin{aligned}\label{eq:c0}
	    C_{i}^{0} =& \sqrt{\frac{\kappa}{N_{c}^{2}-1}} \begin{pmatrix} 0 & 1 \\ 0 & 0 \end{pmatrix} \left(r_{i} + \frac{i p_{i}}{2MT} +\frac{\Delta V_{os}}{4T}r_{i} \right) \\ 
	&+ \sqrt{\kappa} \begin{pmatrix} 0 & 0 \\ 1 & 0 \end{pmatrix} \left(r_{i} + \frac{i p_{i}}{2MT} +\frac{\Delta V_{so}}{4T}r_{i} \right),
	\end{aligned}\\
	&C_{i}^{1} = \sqrt{\frac{\kappa(N_{c}^{2}-4)}{2(N_{c}^{2}-1)}} \begin{pmatrix} 0 & 0 \\ 0 & 1 \end{pmatrix} \left(r_{i} + \frac{i p_{i}}{2MT} \right).\label{eq:c1}
\end{align}
with $\Delta V_{os} = V_o - V_s$ being the difference of the octet and singlet potentials, with $V_s = -C_F \alpha_s/r$ and \mbox{$V_o = C_F \alpha_s/8r$}.  The transport coefficients $\kappa$ and $\gamma$ are given by the following chromoelectric correlators 
\begin{eqnarray}
\kappa &=& \frac{g^{2}}{18} \int_{0}^{\infty} dt \left\langle \left\{ \tilde{E}^{a,i}(t,\mathbf{0}), \tilde{E}^{a,i}(0,\mathbf{0}) \right\} \right\rangle, \\
\gamma &=& -i \frac{g^{2}}{18} \int_{0}^{\infty} dt \left\langle \left[ \tilde{E}^{a,i}(t,\mathbf{0}), \tilde{E}^{a,i}(0,\mathbf{0}) \right] \right\rangle \, .
\end{eqnarray}
Above, $\tilde{E}$ is a chromoelectric field sandwiched by two links in the fundamental representation, i.e.
$\tilde{E}^{a}_i(s, \vec{0}) = \Omega(s)^\dagger E^{a}_i(s, \vec{0}) \Omega(s)$ with
\begin{equation}
    \Omega(s) = \text{exp}\left[  -ig \int_{-\infty}^{s} \text{d}s' A_{0}(s', \vec{0}) \right] .
\end{equation}

The Lindblad equation given in Eq.~(\ref{eq:Lindblad}) together with the collapse operators given in Eqs.~\eqref{eq:c0} and \eqref{eq:c1} describes the evolution of the heavy quarkonium reduced density matrix at NLO $E/T$ \cite{Brambilla:2022ynh}.  For details concerning the derivation of these equations we refer the reader to \cite{Brambilla:2022ynh}.

The transport coefficients $\hat\kappa$ and $\hat\gamma$ are inputs, which can be fixed from direct and indirect lattice measurements.  In Ref.~\cite{Brambilla:2023hkw} it was found that the values of $\hat\kappa=4$ and $\hat\gamma=0$ resulted in a good description of both the ground and excited bottomonium suppression as a function of $N_{\rm part}$ and $p_T$.  These values are consistent with recent lattice extractions which indicate that there is little to no mass shift for $\Upsilon$ states, while the decay widths all bottomonium states are large and increasing functions of $T$ \cite{Larsen:2019bwy,Bala:2021fkm}.  In this work, at all three energies, we assume $\hat\gamma=0$.  At LHC energies we vary $\hat\kappa \in \{3,4\}$ and at RHIC energy we vary it in $\hat\kappa \in \{4,5\}$, since it is expected that $\hat\kappa$ is larger at lower temperature near the QGP phase transition.

\subsection{Real-time numerical solution}

To compute the survival probabilities of various bottomonium states we must solve the NLO Lindblad equation Eq.~\eqref{eq:Lindblad} including effects of quantum jumps (quantum regeneration). For this purpose, we use the quantum trajectories method introduced in Refs.~\cite{Brambilla:2020qwo,Omar:2021kra,Brambilla:2022ynh,Brambilla:2023hkw}.  For all results presented in this work, we used a one-dimensional lattice with $2048$ points and a box size of $L = 40\,\mathrm{GeV}^{-1}$. The temporal step size was taken to be $0.001\,\mathrm{GeV}^{-1}$.  The code used to generate all results presented is available publicly \cite{Omar:2021kra,qtraj-download}.  The input to the code is the temperature experienced by a bottomonium state along its propagation through the QGP.  In the next section we describe how we sample physical bottomonium trajectories and couple them to the background 3D viscous hydrodynamical evolution.

To initialize the real-time quantum evolution, we assume that at $\tau = 0$ fm the wave function is a localized delta function and the system is in the singlet state.  In practice, the initial reduced wave function $u = r \psi$ is given by a Gaussian multiplied by a power of $r$ appropriate for the angular momentum of the state $l$, i.e.,
\begin{equation}
u_{\ell}(t_0) \propto r^{l+1} e^{-r^{2}/(ca_{0})^2},
\end{equation} 
with $u$ normalized to one and $c=0.2$ following earlier works~\cite{Omar:2021kra}. 
We note that observables do not depend significantly on $c$ below this value (see Fig.~6 of Ref.~\cite{Omar:2021kra}).
We evolve the initial wave function using the vacuum Coulomb potential from $\tau = 0$~fm to $\tau_{\rm med}$ = 0.6~fm at which time we turn on the medium interactions.  
We note that this is the same medium initialization time as was used in prior works~\cite{Brambilla:2020qwo,Brambilla:2021wkt,Brambilla:2022ynh}.

\section{Coupling to 3D viscous hydrodynamics}
\label{sec:hydro}

We consider 5.02 TeV, 2.76 TeV, and 200 GeV heavy-ion collisions with the background temperature evolution given by 3+1D quasiparticle anisotropic hydrodynamics (aHydro).  The hydrodynamic parameters were previously tuned to reproduce experimentally observed soft hadron spectra, elliptic flow, and HBT radii  \cite{Alqahtani:2020paa,Alqahtani:2017jwl,Alqahtani:2017tnq,Alqahtani:2017mhy,Almaalol:2018gjh,Alqahtani:2020daq,Alalawi:2021jwn,Alqahtani:2022erx}.

In the cases considered in this work, we use optical Glauber initial conditions specified at $\tau=0.25$ fm/c.  At 5.02 TeV, 2.76 TeV, and 200 GeV best fits of aHydro results to data give initial central temperature of $T_0=$ 630, 600, and 455 MeV, and shear viscosity to entropy density ratios of $\eta/s = 0.179$,  $0.159$, and $0.159$, respectively \cite{Alqahtani:2020paa,Alqahtani:2017jwl,Alqahtani:2017tnq,Almaalol:2018gjh}.  
All other transport coefficients, e.g., the bulk viscosity, are self-consistently determined once $\eta/s$ is fixed.  For details concerning the hydrodynamic simulations we refer to reader to Refs.~\cite{Alqahtani:2020paa,Alqahtani:2017jwl,Alqahtani:2017tnq,Almaalol:2018gjh}.

Hydrodynamic runs were made for a set of impact parameters and the 3+1D Milne-space evolution of the temperature was saved to disk.
In the next section we detail how to convert Cartesian eikonal trajectories to Milne coordinates in order use the generated 3+1D hydrodynamic backgrounds in the computation of the quarkonium survival probability.

\subsection{Bottomonium trajectories in Milne Coordinates}
\label{subsec:milne}

In this work we sample 3D bottomonium trajectories and use the temperature along each trajectory provided by anisotropic hydrodynamics.  Final observables are averaged over a large set of sampled physical (and quantum) trajectories.  In this work we assume that, after their initial production, bottomonium states travel along eikonal trajectories.  Since the hydrodynamic background used is provided in Milne coordinates, we must express spacetime positions using the proper time $\tau$, the transverse coordinates $\vec{x}_\perp$, and the spatial rapidity, $\varsigma$.  Expressed in Cartesian coordinates, we assume that the bottomonium states move at a constant velocity
$\vec{x} = \vec{x}_{0} + \vec{v} (t - t_{0})$,
where $\vec{x}_0$ is the position of the state at $t_0$ and $\vec{v} = \vec{p}_0/E$ with $E = \sqrt{\vec{p}^2 + m^2}$. Transforming to Milne coordinates using
\begin{eqnarray}
t &=& \tau \cosh{\varsigma} \, , \nonumber \\
z &=& \tau \sinh{\varsigma} \, , \label{eqn:space-time}
\end{eqnarray} 
and $v_z= \tanh{y}$, with $y$ being the momentum rapidity, one has
\begin{eqnarray}
\vec{x}_\perp &=& \vec{x}_{\perp,0} + \vec{v}_\perp (\tau \cosh{\varsigma} - \tau_{0} \cosh{\varsigma_{0}}) \, ,  \nonumber \\
\tau \sinh{\varsigma} &=& \tau_{0} \sinh{\varsigma_{0}} + \tanh{y} \, (\tau \cosh{\varsigma} - \tau_{0} \cosh{\varsigma_{0}}) , \;\;\;\;
\end{eqnarray}
where $\varsigma_0$ is the spatial rapidity at $\tau_0$ and $\vec{v}_\perp$ represents the transverse velocity.
These equations allow one to obtain the Milne coordinates of a bottomonium state based on its initially sampled momentum and Milne coordinate position.  We note that the third equation can be solved for $\varsigma$ giving $\varsigma = y + \log({\cal G})$ with 
\begin{equation}
{\cal G} \equiv \sqrt{1 + \frac{\sinh^2(\varsigma_0-y)}{\bar\tau^2}} + \frac{\sinh(\varsigma_0-y)}{\bar\tau}  \, ,
\end{equation}
and $\bar\tau \equiv \tau/\tau_0$.  We note that in the limit $\bar\tau \rightarrow \infty$, one has ${\cal G} \rightarrow 1$, and in the limit $\bar\tau \rightarrow 1$, one has ${\cal G} \rightarrow \exp(\varsigma_0-y)$. In the limit that all production occurs at $\tau_0 \rightarrow 0$, one has $\varsigma=\varsigma_0 = y$ at all times.

For bottomonium production we Monte-Carlo sample the initial transverse position from the binary collision overlap profile of the two gold nuclei.  We sample the initial momentum rapidity, $\varsigma_0 = y_0$, using a Gaussian width of $2.3$.  This value was set using fits to experimental data for the momentum rapidity dependence of bottomonium production in $pp$ collisions~\cite{Woehri:2014}.  This rapidity width was also checked using Pythia 8 $\Upsilon(1S)$ 8 TeV pp production results and we found that the width extracted using Pythia 8 is consistent with experimental measurements of the rapidity dependence of $\Upsilon(1S)$ production. The initial transverse momenta are sampled from a $1/E_T^4$ distribution where $E_T = \sqrt{p_T^2 + \langle M \rangle^2}$ and \mbox{$ \langle M \rangle = 10$ GeV}. We then sample the azimuthal angle $\phi$ of the produced bottomonium wave packet uniformly in the $[0,2\pi)$.

\section{Excited state feed down}
\label{sec:feeddown}

For all cases presented in this work we apply the same late-time excited state feed down. Feed down is accounted for using a feed down matrix $F$ which relates the experimentally observed and direct production cross sections, $\vec{\sigma}_{\text{exp}} = F \vec{\sigma}_{\text{direct}}$.  In this relation, $F$ is a square matrix whose values are fixed by the experimentally extracted branching fractions of the bottomonium excited states into lower lying states.
The states considered are the $\Upsilon(1S),\,$ $\Upsilon(2S),\,$ $\chi_{b0}(1P),\,$ $\chi_{b1}(1P),\,$ $\chi_{b2}(1P),\,$ $\Upsilon(3S),\,$ $\chi_{b0}(2P),\,$ $\chi_{b1}(2P),\,$ and $\chi_{b2}(2P)$.
The corresponding entries in the feed down matrix $F$ are
\begin{equation}
	F_{ij} = \left\{ \begin{matrix}
		\; \text{branching fraction $j$ to $i$},\; & \text{for } i < j , \\
		1, & \text{for } i = j , \\
		0, & \text{for } i > j ,
		\end{matrix} \right.
\end{equation}
where the branching fractions are taken from the Particle Data Group~\cite{Zyla:2020zbs} (see also Eq.~(6.4) of Ref.~\cite{Brambilla:2020qwo}).

We compute the nuclear modification factor $R_{AA}^i$ for a state $i$ using
\be
R^{i}_{AA}(c,p_T,\phi) = \frac{\left(F \cdot S(c,p_T,\phi) \cdot \vec{\sigma}_{\text{direct}}\right)^{i}}{\vec{\sigma}_{\text{exp}}^{i}} \, ,
\label{eq:feeddown}
\ee
where $S(c,p_T,\phi)$ is the survival probability computed from the real-time quantum evolution; $c$ is the centrality class, $p_T$ is transverse momentum, and $\phi$ is azimuthal angle.

For RHIC energies, we use the following integrated pp cross sections, $\vec{\sigma}_{\text{exp}} = \{2.35$, 0.775, 0.152, 0.559, 0.657, 0.277, 0.133, 0.490, $0.577\}$ nb.  These values are consistent with measurements by the STAR collaboration at $\sqrt{s} = $ 200 GeV \cite{Ye:2017fwv} and with LHC observations when scaled in collision energy \cite{Sirunyan:2018nsz,Aaij:2014caa,Brambilla:2020qwo}.  
For LHC 5.02 TeV energy collisions, the pp cross sections used were $\vec{\sigma}_{\text{exp}}=\{57.6$, 19, 3.72, 13.69, 16.1, 6.8, 3.27, 12.0, $14.15\}$ nb.
These were obtained from experimental measurements presented in Refs.~\cite{Sirunyan:2018nsz,Aaij:2014caa} as explained in Sec.~6.4 of Ref.~\cite{Brambilla:2020qwo}.  For LHC 2.76 TeV energy collisions, we use scaled 5.02 TeV cross-sections.

\begin{figure}[t]
\includegraphics[width=0.97\linewidth]{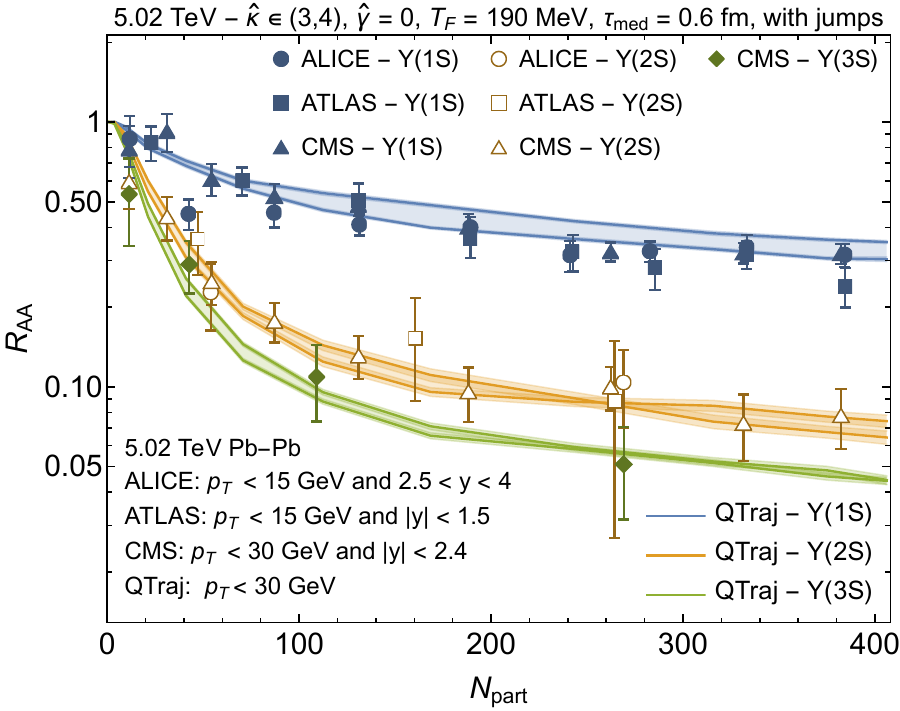}
\caption{QTraj predictions for $R_{AA}$ as a function of $N_{\rm part}$ in 5.02 TeV Pb-Pb collisions. The experimental data shown are from the ALICE \cite{Acharya:2020kls}, ATLAS \cite{ATLAS:2022xso}, and CMS \cite{CMS:2018zza,CMS:2023lfu} collaborations. The lighter shaded bands indicate the uncertainty associated with the choice of $\hat\kappa$ and the darker shaded bands on each boundary indicate the statistical uncertainty associated with the average over bottomonium trajectories.}
\label{fig:raavsnpart-lhc-3d}
\end{figure}

\begin{figure}[t!]
\includegraphics[width=0.97\linewidth]{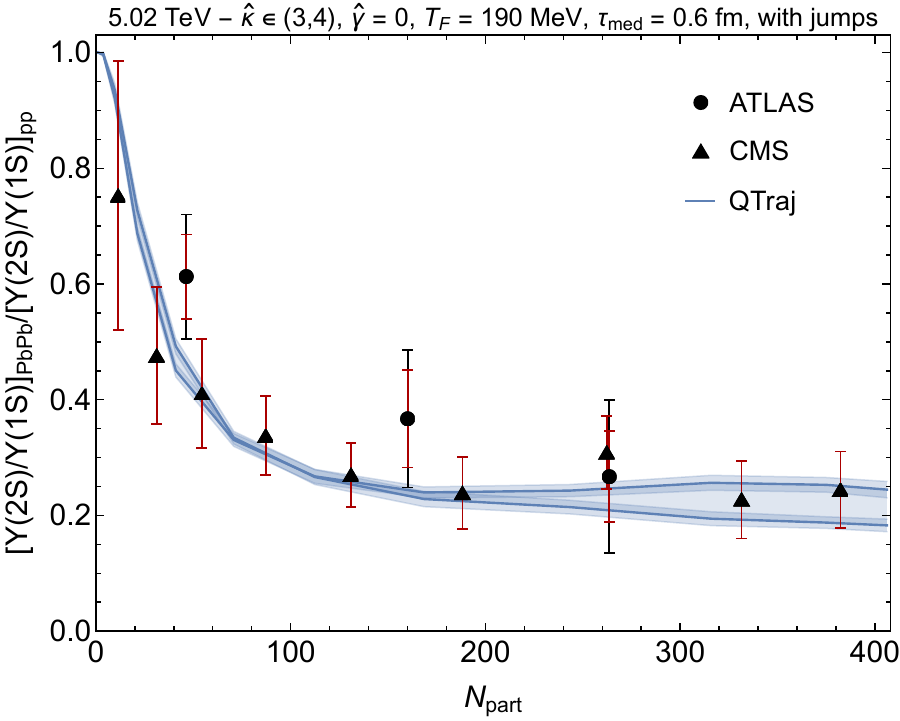}
\caption{QTraj predictions for the 2S to 1S double ratio as a function of $N_{\rm part}$ in 5.02 TeV Pb-Pb collisions. Line styles and experimental data sources are the same as in Fig.~\ref{fig:raavsnpart-lhc-3d}.}
\label{fig:ratio21vsnpart-lhc-3d}
\end{figure}

\section{Results}
\label{sec:results}

For the results presented below we sampled physical trajectories distributed evenly in following centrality classes (0{-}0.05), (0.05{-}0.1), (0.1{-}0.2), (0.2{-}0.3), (0.3{-}0.4), (0.4{-}0.5), (0.5{-}0.6), (0.6{-}0.7), (0.7{-}0.8), (0.8{-}0.90), and (0.9{-}1). In each of these centrality classes, at $\tau=0^+$, the initial positions and momenta of the bottomonia were sampled as described in Sec.~\ref{subsec:milne}. Along each physical trajectory we solved the Lindblad equation \eqref{eq:Lindblad} using the quantum trajectories method.  The open-source code used for this is called QTraj 2.0 \cite{qtraj-download}.  Finally, feed down of excited states was taken into account as described in Sec.~\ref{sec:feeddown}.

\subsection{5 TeV Pb-Pb collisions}

We begin by comparing the NLO OQS+pNRQCD predictions with $\sqrt{s_{NN}} = 5.02$ TeV Pb-Pb data from the ALICE \cite{Acharya:2020kls}, ATLAS \cite{ATLAS:2022xso}, and CMS \cite{CMS:2018zza,CMS:2023lfu} collaborations.  The three different experimental collaborations had rapidity coverages of $2.5 < y < 4$, $|y| < 1.5$, and $|y| < 2.4$, respectively.  For the results at this collision energy we sampled approximately 200k physical trajectories and 40 quantum trajectories per physical trajectory.

In Fig.~\ref{fig:raavsnpart-lhc-3d} we present $R_{AA}[\Upsilon(1S)]$, $R_{AA}[\Upsilon(2S)]$ , and $R_{AA}[\Upsilon(3S)]$ as a function of the number of participants $N_{\rm part}$.
For the QTraj results, the lighter shaded bands indicate the uncertainty associated with the choice of $\hat\kappa$ and the darker shaded bands around each line indicate the statistical uncertainty associated with the average over bottomonium trajectories.  We have plotted the comparison on a logarithmic scale in order to better see the details of the excited state suppression.  Fig.~\ref{fig:raavsnpart-lhc-3d} demonstrates that NLO OQS+pNRQCD is able to well-describe the $N_{\rm part}$ (centrality) dependence of all three observed bottomonium states.  When compared to results obtained with zero rapidity bottomonium states in Ref.~\cite{Brambilla:2023hkw}, we see that $R_{AA}[\Upsilon(1S)]$ is very similar, however, the agreement of both $R_{AA}[\Upsilon(2S)]$ and $R_{AA}[\Upsilon(3S)]$ with experimental data is improved at $N_{\rm part} \lesssim 200$.

Next we turn to the ``double ratio'' obtained by taking the ratio of $\Upsilon(2S)$ to $\Upsilon(1S)$ production in Pb-Pb and pp collisions, defined as $(n[2S]/n[1S])_{AA}/(n[2S]/n[1S])_{pp}$.  Such double ratios have been shown to be sensitive to quantum regeneration of bottomonium excited states in the QGP \cite{Brambilla:2023hkw}.  We present our result for the 2S to 1S double ratio in Fig.~\ref{fig:ratio21vsnpart-lhc-3d} and compare our predictions with experimental data from the ATLAS and CMS collaborations.  As this figure demonstrates, there is only a small variation observed when varying the heavy quark transport coefficient $\hat\kappa$.  This is due to the fact that increasing $\hat\kappa$ increases the suppression of both the ground and excited states, with this effect largely canceling in the ratio.  Our results for the 2S to 1S double ratio show good agreement with the available experimental data and, when compared to prior work using the OQS+pNRQCD framework \cite{Brambilla:2023hkw}, we find that allowing for bottomonia states with finite rapidity improves agreement with experimental data for $N_{\rm part} \lesssim 200$.

\begin{figure}[t!]
\includegraphics[width=0.97\linewidth]{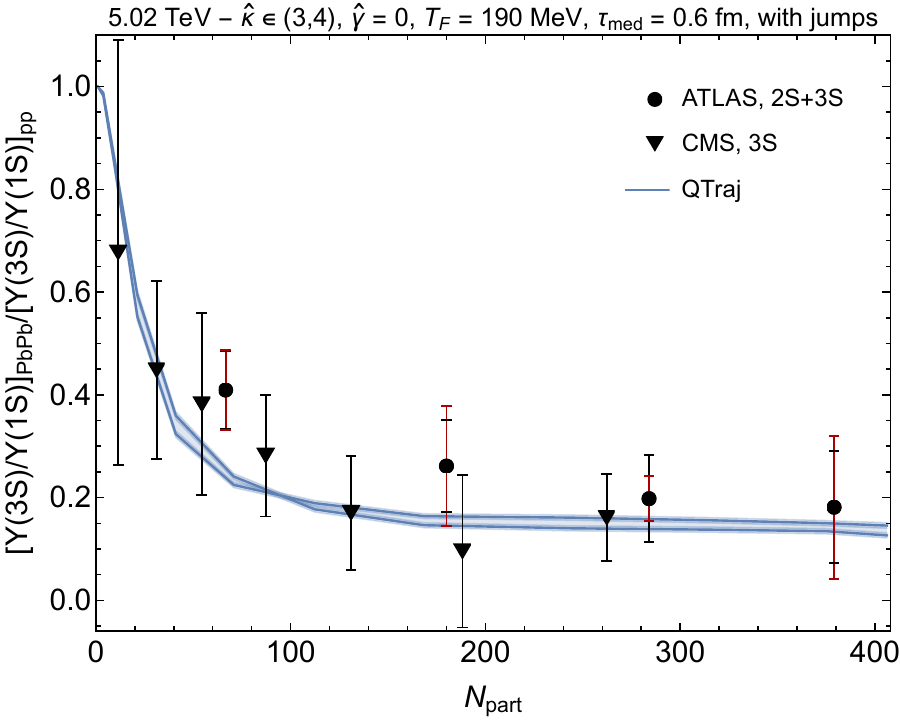}
\caption{QTraj predictions for the 3S to 1S double ratio as a function of $N_{\rm part}$ in 5.02 TeV Pb-Pb collisions. Line styles and experimental data sources are the same as in Fig.~\ref{fig:raavsnpart-lhc-3d}.}
\label{fig:ratio31vsnpart-lhc-3d}
\end{figure}

Similarly, one can see from Fig.~\ref{fig:ratio31vsnpart-lhc-3d} that OQS+pNRQCD results for the 3S to 1S double ratio are in very good agreement with the experimental data from the CMS collaboration.  We note that the ATLAS results shown are for the integrated 2S+3S double ratio and so is expected to be between the true 3S to 1S and 2S to 1S ratios.  With regards to the OQS+pNRQCD prediction, we see that the result obtained is nearly independent of assumed value of the $\hat\kappa$, allowing for a quite constrained theoretical prediction.

\begin{figure}[t!]
\includegraphics[width=0.97\linewidth]{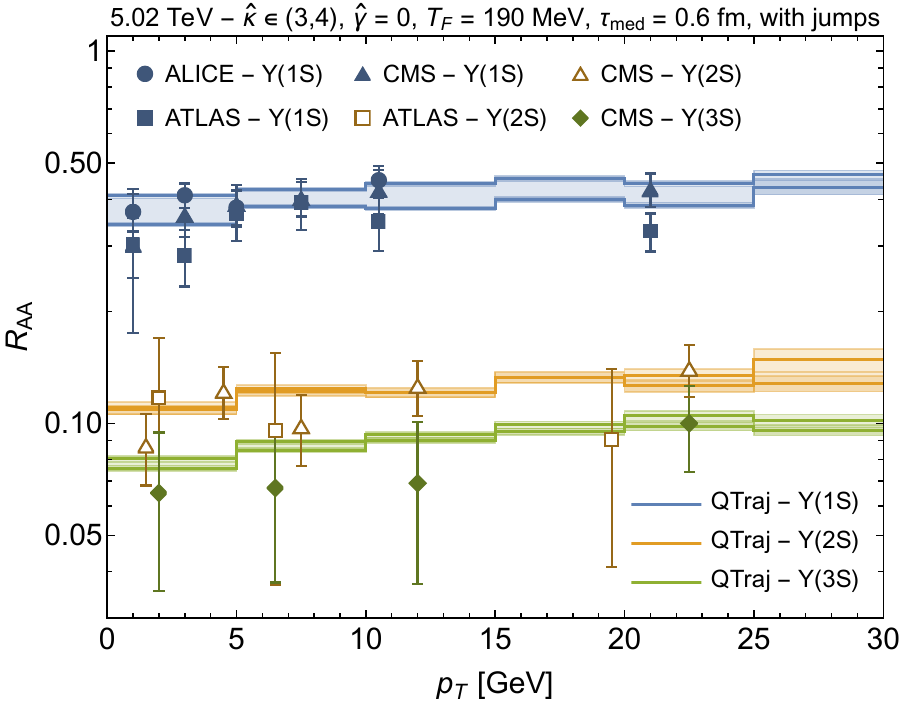}
\caption{QTraj predictions for $R_{AA}$ as a function of $p_T$ in 5.02 TeV Pb-Pb collisions. Line styles and experimental data sources are the same as in Fig.~\ref{fig:raavsnpart-lhc-3d}.}
\label{fig:raavspt-lhc-3d}
\end{figure}

In Fig.~\ref{fig:raavspt-lhc-3d} we present our predictions for $R_{AA}[\Upsilon(1S)]$, $R_{AA}[\Upsilon(2S)]$ , and $R_{AA}[\Upsilon(3S)]$ as a function of transverse momentum $p_T$.  As in the previous figures the light bands indicate the variation with $\hat\kappa$ while the darker bands indicate the statistical uncertainty associated with the average over trajectories.  In order to increase statistics, we binned $p_T$ in 5 GeV intervals and in order to better resolve the suppression of the excited states, we have used a logarithmic scale on the vertical axis.  As can be seen from this Figure, the OQS+pNRQCD approach predicts a rather flat distribution, with the $p_T$ dependence coming solely from the mean lifetime of the state in the QGP.  This is consistent with experimental observations from the ALICE, ATLAS, and CMS collaborations.

\begin{figure}[t!]
\includegraphics[width=0.97\linewidth]{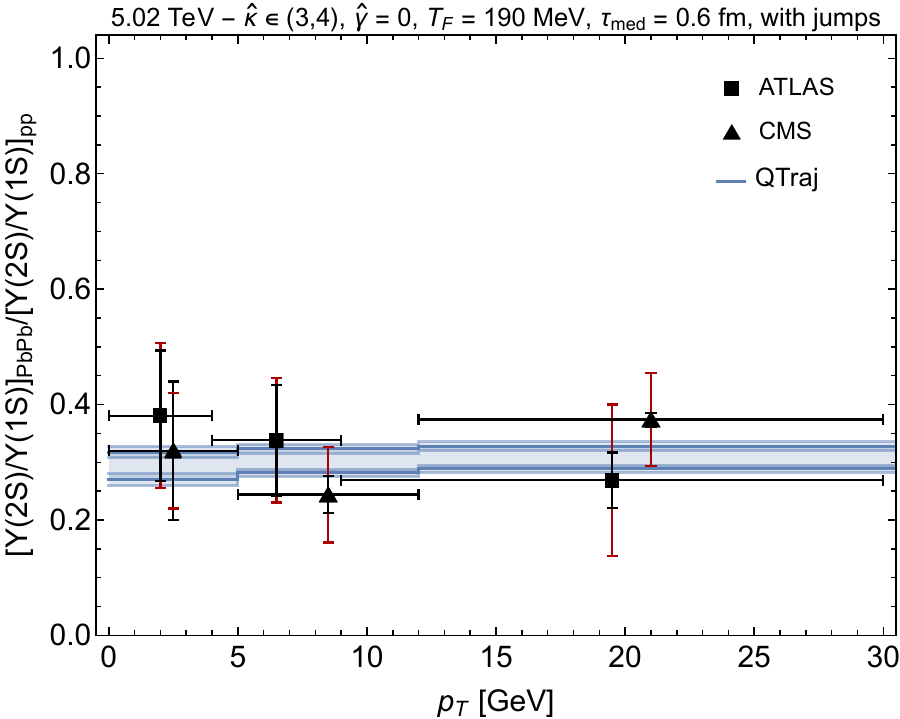}
\caption{QTraj predictions for the 2S to 1S double ratio as a function of $p_T$ in 5.02 TeV Pb-Pb collisions. Line styles and experimental data sources are the same as in Fig.~\ref{fig:raavsnpart-lhc-3d}.}
\label{fig:ratio21vspt-lhc-3d}
\end{figure}

\begin{figure}[t!]
\includegraphics[width=0.97\linewidth]{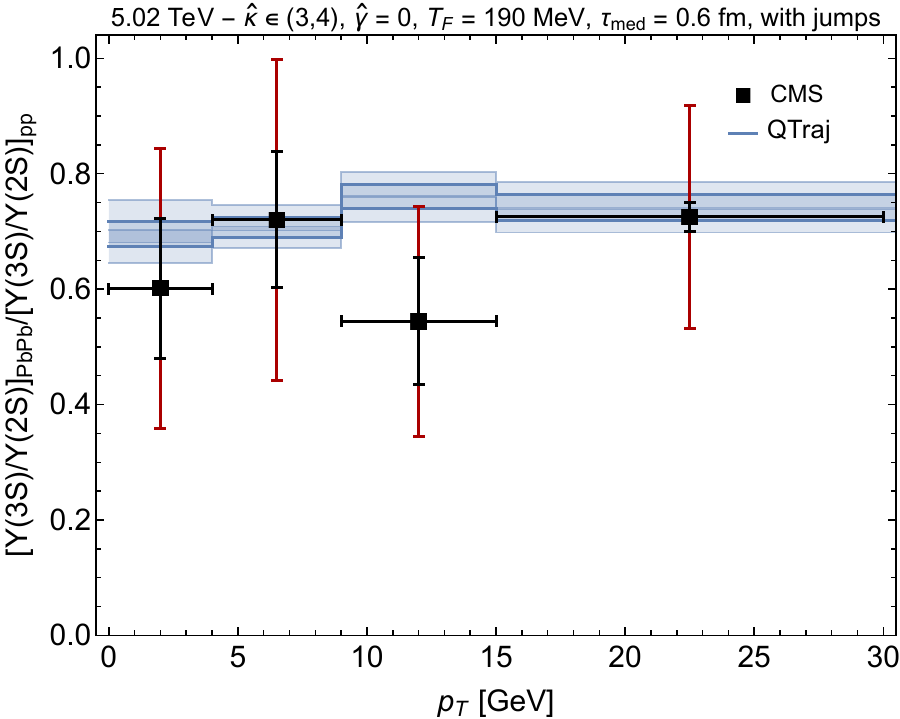}
\caption{QTraj predictions for the 3S to 2S double ratio as a function of $p_T$ in 5.02 TeV Pb-Pb collisions. Line styles and experimental data sources are the same as in Fig.~\ref{fig:raavsnpart-lhc-3d}.}
\label{fig:ratio32vspt-lhc-3d}
\end{figure}

\begin{figure}[t!]
\includegraphics[width=0.97\linewidth]{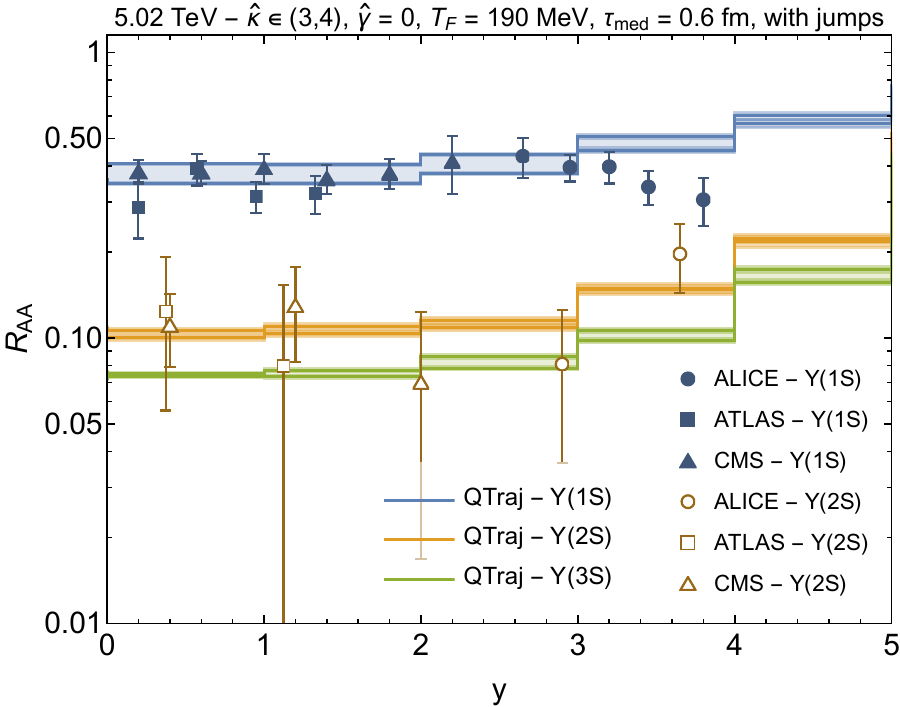}
\caption{QTraj predictions for $R_{AA}$ as a function of $y$ in 5.02 TeV Pb-Pb collisions. Line styles and experimental data sources are the same as in Fig.~\ref{fig:raavsnpart-lhc-3d}.}
\label{fig:raavsy-lhc-3d}
\end{figure}

The 2S to 1S double ratio as a function of transverse momentum has been reported by both the ATLAS and CMS collaborations and is shown in Fig.~\ref{fig:ratio21vspt-lhc-3d}.  For this figure we have used three $p_T$ bins for the QTraj results:  (0,5), (5,12), and $>$ 12 GeV.  As we can see from this Figure, the 2S to 1S double ratio shows a very weak dependence on $p_T$ and our predictions are consistent with experimental observations.  Similar agreement is found when considering the $p_T$ dependence of 3S to 2S double ratio shown in Fig.~\ref{fig:ratio32vspt-lhc-3d}.  In this Figure we compare our predictions to recent data reported by the CMS collaboration \cite{CMS:2023lfu}.  Once again we see that the double ratio predicted by the OQS+pNRQCD approach does not depend strongly on transverse momentum.

Finally, in Fig.~\ref{fig:raavsy-lhc-3d} we present our predictions for $R_{AA}[\Upsilon(1S)]$, $R_{AA}[\Upsilon(2S)]$ , and $R_{AA}[\Upsilon(3S)]$ as a function of rapidity $y$ for $0 < y < 5$.  As this figure demonstrates the OQS+pNRQCD approach predicts that $R_{AA}$ is quite flat at central rapidities and increases as one approaches forward or backward rapidities due to the fact that the temperature is maximal at central rapidities and falls off as one goes to forward/backward rapidity.  We note that there is some tension with the ALICE data collected in the interval $2.4 < y < 4$ since there are indications from the ALICE data that $R_{AA}$ decreases at large rapidity.  It will be very interesting to get high statistics experimental results for both ground and excited state suppression in this rapidity interval to see if this tension persists.  One possible explanation for the discrepancy is that this is related to nuclear parton distribution function (nPDF) effects, which need to be taken account at large rapidities \cite{Vogt:2010aa}.

\begin{figure}[t!]
\includegraphics[width=0.97\linewidth]{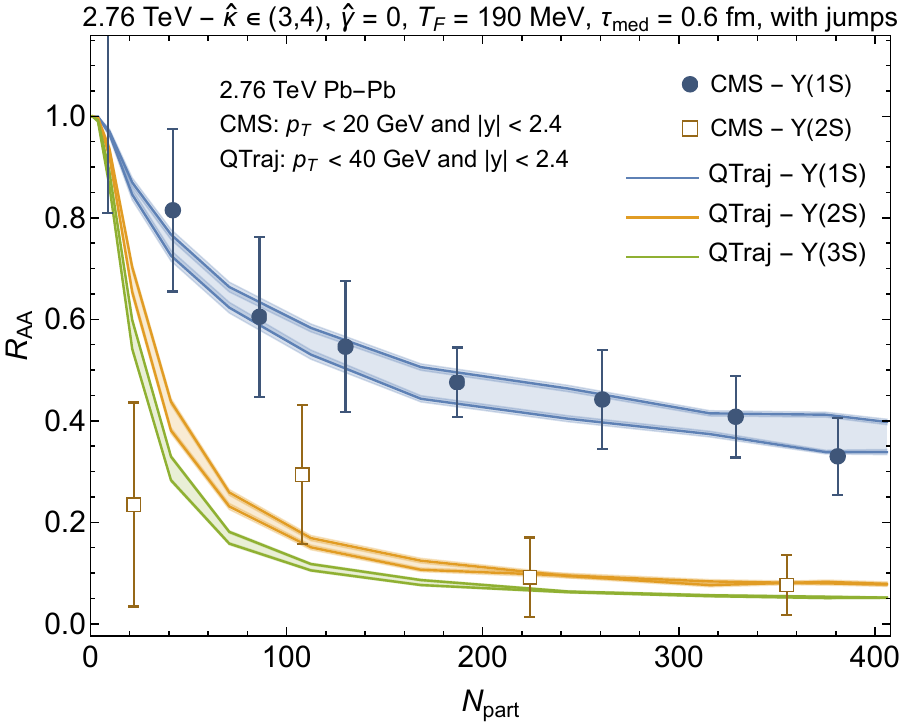}
\caption{QTraj predictions for $R_{AA}$ as a function of $N_{\rm part}$ in 2.76 TeV Pb-Pb collisions.  The experimental data shown are from the ALICE \cite{ALICE:2014wnc} and CMS \cite{CMS:2016rpc} collaborations.}
\label{fig:raavsnpart-lhc-2.76-3d}
\end{figure}

\begin{figure}[t!]
\includegraphics[width=0.97\linewidth]{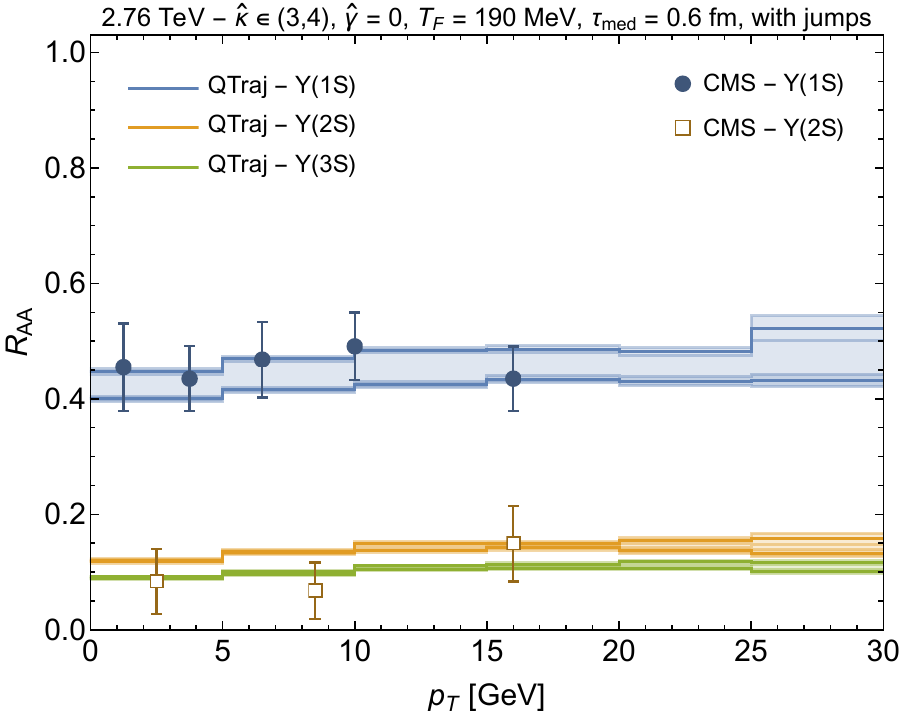}
\caption{QTraj predictions for $R_{AA}$ as a function of $p_T$ in 2.76 TeV Pb-Pb collisions. The experimental data sources are the same as in Fig.~\ref{fig:raavsnpart-lhc-2.76-3d}.}
\label{fig:raavspt-lhc-2.76-3d}
\end{figure}

\begin{figure}[t!]
\includegraphics[width=0.97\linewidth]{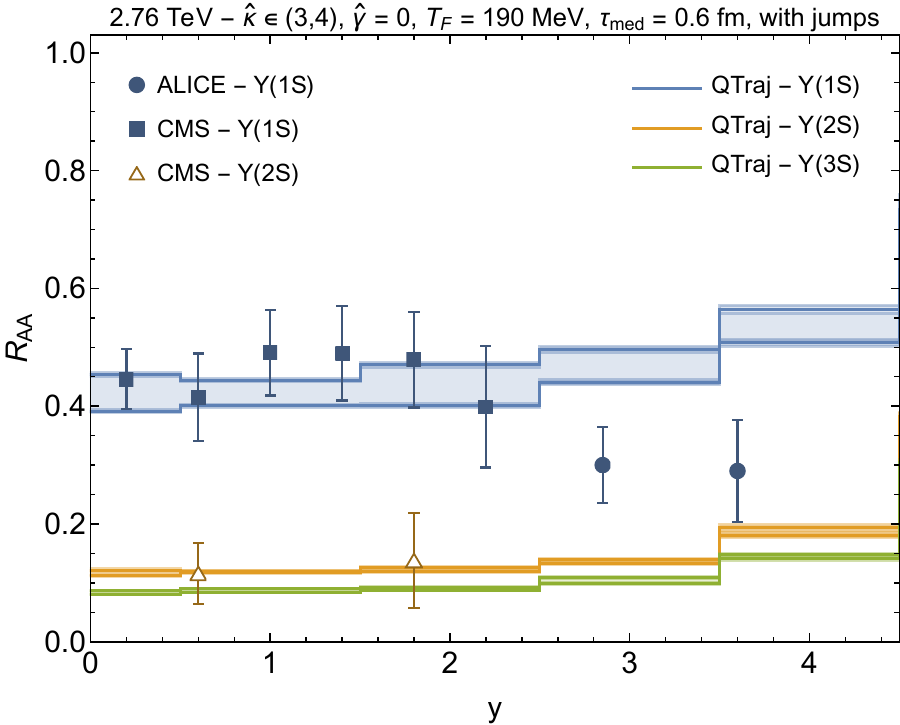}
\caption{QTraj predictions for $R_{AA}$ as a function of $y$ in 2.76 TeV Pb-Pb collisions. The experimental data sources are the same as in Fig.~\ref{fig:raavsnpart-lhc-2.76-3d}.}
\label{fig:raavsy-lhc-2.76-3d}
\end{figure}
\subsection{2.76 TeV Pb-Pb collisions}

Next we consider 2.76 TeV Pb-Pb collisions in order to test the predictive power of the OQS+pNRQCD framework.  We note that, when going from $\sqrt{s_{NN}} =$ 5.12 TeV to 2.76 TeV, the initial central temperature decreases by only 30 MeV \cite{Alqahtani:2017tnq,Alqahtani:2020paa}, but even this small difference causes a measurable decrease in $R_{AA}$ when increasing the beam energy.  Besides changing the hydrodynamic background we did not change any other of the parameters, e.g. $\hat\kappa$ $\hat\gamma$, $\tau_{\rm med}$, etc. We will compare the predictions of the OQS+pNRQCD approach with data from the ALICE \cite{ALICE:2014wnc} and CMS \cite{CMS:2016rpc} collaborations.
For the QTraj results at this collision energy we sampled approximately 275k physical trajectories and 40 quantum trajectories per physical trajectory.

In Figs.~\ref{fig:raavsnpart-lhc-2.76-3d} and \ref{fig:raavspt-lhc-2.76-3d} we present $R_{AA}[\Upsilon(1S)]$, $R_{AA}[\Upsilon(2S)]$, and $R_{AA}[\Upsilon(3S)]$ as a function of the number of participants $N_{\rm part}$ and transverse momentum $p_T$, respectively.  As in the case of 5.02 TeV collisions, we varied $\hat\kappa \in (3,4)$ while holding $\hat\gamma=0$ \cite{Brambilla:2023hkw}.  The resulting OQS+pNRQCD predictions show good agreement with the experimental data both as a function of $N_{\rm part}$ and $p_T$.  

Finally, turning to Fig.~\ref{fig:raavsy-lhc-2.76-3d} we present the OQS+pNRQCD predictions for the rapidity dependence of $R_{AA}$ for all three s-wave states.  As in the case of 5.02 TeV collisions, we observe that the OQS+pNRQCD predictions agree well with the experimental data for $y \lesssim 2$, while above this the ALICE collaboration's data indicate that $R_{AA}$ is smaller at large rapidity.  Once again, the most likely explanation for this is nPDF effects namely gluon shadowing, which may be large at large rapidity \cite{Vogt:2010aa}.

\begin{figure}[t!]
\includegraphics[width=0.97\linewidth]{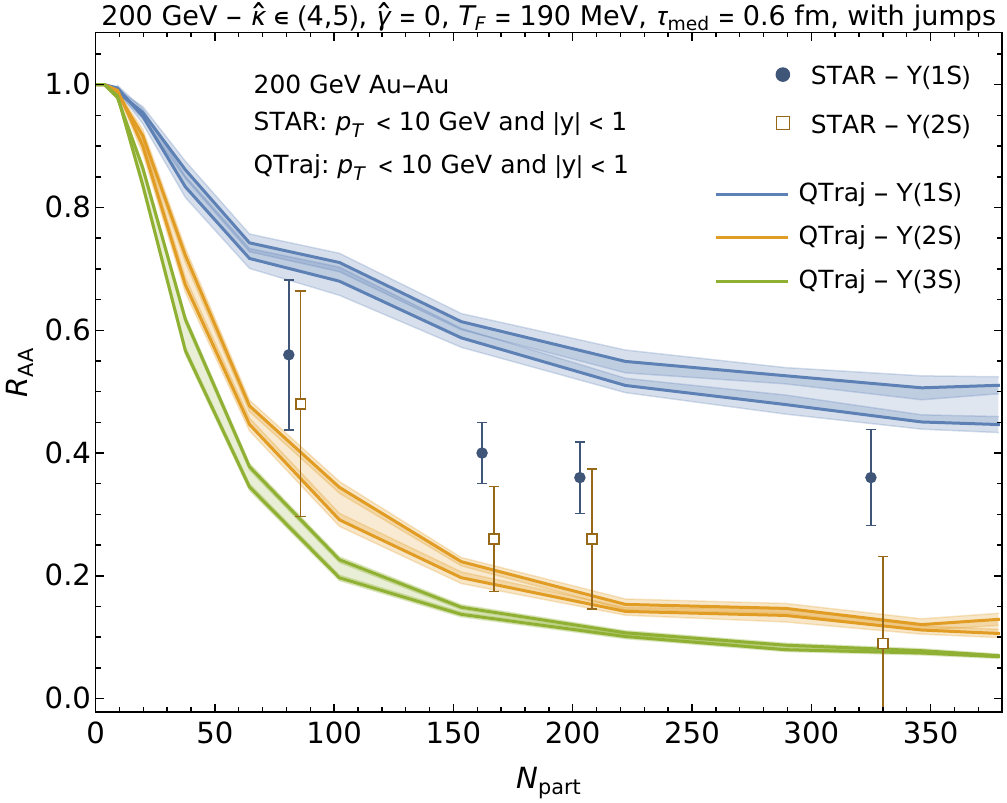}
\caption{QTraj predictions for $R_{AA}$ as a function of $N_{\rm part}$ in 200 GeV Au-Au collisions.  The experimental data shown are from the STAR collaboration \cite{STAR:2022rpk}.}
\label{fig:raavsnpart-rhic-3d}
\end{figure}

\begin{figure}[t!]
\includegraphics[width=0.97\linewidth]{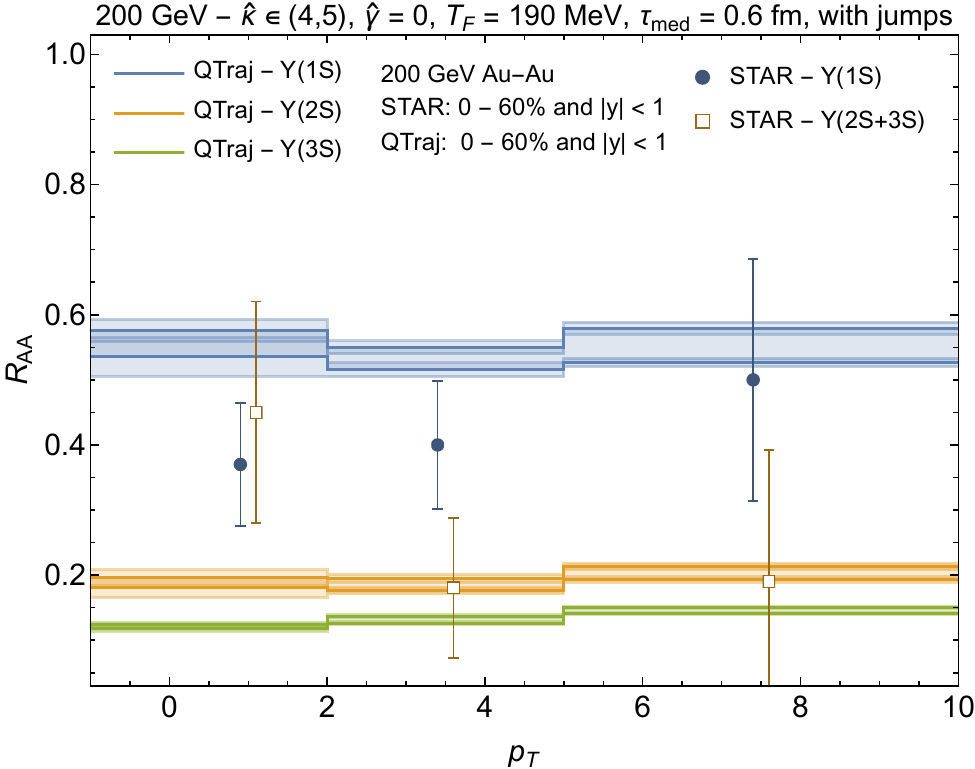}
\caption{QTraj predictions for $R_{AA}$ as a function of $p_T$ in 200 GeV Au-Au collisions. The experimental data sources are the same as in Fig.~\ref{fig:raavsnpart-rhic-3d}.}
\label{fig:raavspt-rhic-3d}
\end{figure}

\begin{figure}[t!]
\includegraphics[width=0.97\linewidth]{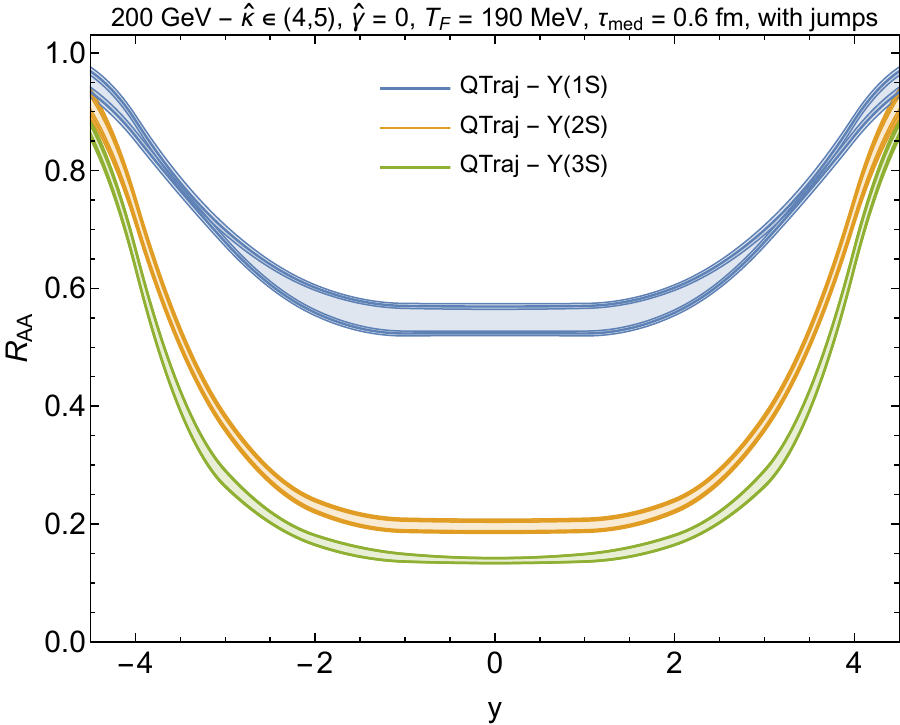}
\caption{QTraj predictions for $R_{AA}$ as a function of $y$ in 200 GeV Au-Au collisions. The experimental data sources are the same as in Fig.~\ref{fig:raavsnpart-rhic-3d}.}
\label{fig:raavsy-rhic-3d}
\end{figure}

\subsection{200 GeV Au-Au collisions}

Next we consider $\sqrt{s_{NN}} = 200$ GeV Au-Au collisions and compare our predictions to findings recently reported by the STAR collaboration \cite{STAR:2022rpk}.  The STAR collaboration has a rapidity coverage of $|y|<1$ and they report results for transverse momentum $p_T < 10$ GeV.  For the QTraj results at this collision energy we sampled approximately 500k physical trajectories and 40 quantum trajectories per physical trajectory.

Our results are presented in Figs.~\ref{fig:raavsnpart-rhic-3d}-\ref{fig:raavsy-rhic-3d} as a function of $N_{\rm part}$, $p_T$, and $y$.  Due to the fact that it is anticipated that $\hat\kappa$ is larger at lower temperature we have adjusted the range of variation of $\hat\kappa$ to be within $4 \leq \hat\kappa \leq 5$ while still holding $\hat\gamma$ fixed to be zero.
As can be seen from Figs~\ref{fig:raavsnpart-rhic-3d} and \ref{fig:raavspt-rhic-3d} the NLO  OQS+pNRQCD framework seems to underestimate the amount of $\Upsilon(1S)$ suppression observed by the STAR collaboration, while it seems to slightly over predict the amount of $\Upsilon(2S)$ observed.  One possible explanation of this discrepancy is that cold nuclear matter effects are larger at RHIC energies than at LHC energies.  However, at RHIC collision energies one can have enhancement from nPDF effects~\footnote{We have observed such an enhancement by running Pythia 8 and comparing $pp$ and $AA$.}, so this would leave energy loss or comover interactions.  One way to understand what is occurring is to have measurements of the suppression as a function of rapidity $y$.  In Fig.~\ref{fig:raavsy-rhic-3d} we present our predictions for $R_{AA}$ as a function of $y$.  It would be very interesting to compare either STAR or forthcoming sPHENIX data to this prediction in particular.

\section{Conclusions}
\label{sec:conclusions}

In this paper we considered the suppression of bottomonium at three different collision energies.  We made use of a state-of-the-art quantum description based on the application of open quantum system methods to the pNRQCD Lagrangian.  The resulting Lindblad equation for the evolution of the heavy-quark reduced density matrix is accurate to next-to-leading (NLO) order to the binding energy $E$ over the temperature $T$ \cite{Brambilla:2023hkw}, which is important for extending the treatment to lower temperatures.  Compared to prior works we have included the rapidity dependence of the produced bottomonia states, allowing them to propagate both transversely and longitudinally with respect to the beam axis.  For this purpose, we made use of 3+1D hydrodynamical backgrounds provided within the anisotropic hydrodynamics framework, the initial conditions of which were independently tuned to reproduce a large set of soft hadron observables, such as identified hadron multiplicities, flow, etc.

We propagated the sampled bottomonia states through the QGP, sampling the temperature along each trajectory as input to the Lindblad equation for the evolution of the heavy-quark reduced density matrix.  The numerically solved next-to-leading-order Lindblad equation includes the effects of {\em quantum jumps} between different color/angular momentum states, which has been shown previously to allow for quantitative understanding of the suppression of both the ground and excited bottmonium states \cite{Brambilla:2023hkw}.  We find that, at LHC energies, including the rapidity dependence of bottomonium production improves agreement with the data for $R_{AA}$ as a function of $N_{\rm part}$ at $N_{\rm part} \lesssim 200$.  Compared to prior work \cite{Brambilla:2023hkw} we are able to also predict the rapidity dependence of $R_{AA}$, finding excellent agreement with experimental data for $|y| \lesssim 3$, while there seems to be some tension with ALICE collaboration data at large rapidity (see, e.g., Figs.~\ref{fig:raavsy-lhc-3d} and \ref{fig:raavsy-lhc-2.76-3d}).

At RHIC energies, we observed much poorer agreement between model predictions and experimental data recently reported by the STAR collaboration \cite{STAR:2022rpk}.  At these energies the NLO OQS+pNRQCD framework is not able to simultaneously describe 1S and 2S suppression.  Although the predictions for $\Upsilon(2S)$ suppression are consistent with experimental observations within uncertainties, we under predict the amount of $\Upsilon(1S)$ suppression when compared to STAR data.  One might expect that the additional suppression could come from nuclear parton distribution effects (nPDF) and cold nuclear matter effects such as coherent energy loss could explain this.  Future work will focus on including these effects at both RHIC and LHC energies in order to have a more comprehensive understanding of experimental observations.  

Finally, we mention that another possible cause for the poorer agreement with experimental data at RHIC energies is that the temperature is lower and hence the expansion in the binding energy over temperature $E/T$ is becoming more suspect.  This perhaps motivates exploring what happens at next-to-next-to-leading order in $E/T$ or developing a hybrid framework in which the optical limit of the Lindblad equation is used at low temperatures.  We plan to investigate both possibilities in future work.

{\bf Acknowledgments} -- This work was supported by U.S. Department of Energy Award No.~DE-SC0013470.

\bibliographystyle{apsrev4-1}
\bibliography{qtraj3d-inspire,qtraj3d-notinspire}

\end{document}